\def\arcm{\hbox{$^{\prime}$}}

\def\arcs{\hbox{$^{\prime\prime}$}}
\def\farcs{\hbox{$,\!\!^{\prime\prime}$}}
\def\magarcsec{$\rm{mag\,arcsec^{-2}}$}
\newcommand{\Lower}[1]{\smash{\lower 1.5ex \hbox{#1}}}
\newenvironment{mptbl}{\begin{center}}{\end{center}}
\newenvironment{minipagetbl}[1]
{\begin{center}\begin{minipage}{#1}
 \begin{mptbl}}%
{\vspace{-.1in} \end{mptbl} \end{minipage} \end{center}}
\documentclass{aa}
\usepackage{graphicx}
\usepackage{natbib}
\bibpunct{(}{)}{;}{a}{}{,}
\usepackage{textcomp}
\usepackage{txfonts}
\newfont{\Bb}{msbm10}
\begin{document}
   \title{An optical search for Low Surface Brightness Galaxies in the
   Arecibo HI Strip Survey}

   \author{C. Trachternach,  D. J. Bomans, L. Haberzettl,
          \and
          R.-J. Dettmar 
          }

   \offprints{C. Trachternach}

   \institute{Astronomisches Institut, Ruhr-Universit\"at Bochum,
              Universit\"atstra\ss{}e 150, 44780 Bochum, Germany\\
              \email{trachter@astro.rub.de}
             }

   \date{Received <date>/ Accepted 18/07/2006}

\abstract{}
{
In order to estimate the contribution of low surface brightness (LSB) galaxies
to the local ($\rm{z\le 0.1}$) galaxy number density, we performed an optical search for LSB candidates in a 15.5 deg${^{2}}$  part of the
region covered by the 65 deg${^{2}}$ blind Arecibo HI Strip Survey (AHISS).
} 
{
Object
detection and galaxy profile fitting were done with analytical
algorithms. The detection efficiency and the selection effects were
evaluated using large samples of artificial galaxies.}
{
Our final
catalogue is diameter-limited and contains 306 galaxies with diameters
$>18\arcs$ at the limiting surface brightness of
$\mu_{B,lim}=25.2\pm0.31\,$\magarcsec.  Of these 306 galaxies, 148 were
not catalogued previously. Our results indicate that low surface
brightness galaxies contribute at
least to 30\,\% to the local galaxy number density.}
{
Without additional
distance information, choosing the limiting diameter and the surface
brightness at which the diameter is measured is crucial.
Depending on these choices, diameter-limited optical catalogues are either
biased against LSB galaxies, or contaminated with cosmologically dimmed high
surface brightness galaxies, which
affects the implied surface brightness distribution. The comparison to
the AHISS showed that although optical surveys detect more galaxies per
deg$^{2}$ than HI surveys, their drawback is the need
for spectroscopic follow up observations to derive distances. Blind HI
surveys have no diameter limits, but tend to miss gas-poor galaxies and all
galaxies which lie outside their redshift limits.
HI and optical surveys thus provide complementary information and
sample different parts of the LSB galaxy population. }

   
\authorrunning{Trachternach et al.}
 \titlerunning{An Optical Search for LSB Galaxies in the AHISS}
   \maketitle
%

\section{Introduction}

The volume density of low surface brightness (LSB) galaxies has been
underestimated for a long time, as they are quite
underrepresented in many earlier optical catalogues due to strong selection
effects against their detection \citep[a detailed description of these effects
can be found in e.g.,][]{impey-1997-review}.
These selection effects resulted in the Freeman law \citep{freeman-1970} 
indicating that spiral galaxies have a typical inclination corrected central disc surface brightness 
with a relatively small dispersion (${\rm\mu_{B,corr}=(21.65\pm0.3)\,mag\,arcsec^{-2}}$).
\cite{disney-1976} suggested that there could be a large population of galaxies
which was undiscovered by most of the optical surveys due to
selection effects.
Subsequent studies showed that there is indeed a large population of
galaxies with central surface brightnesses much fainter than the Freeman value
\citep[e.g.,][]{impey-1988, schombert-1988}.
More recent estimates were able to show that LSB galaxies account for a significant fraction of the total
galaxy numbers \citep[see e.g.,][]{mcgaugh-1995, mcgaugh-1996,
impey-1997-review, bothun-1997, oneil-bothun-2000}. 
\cite{minchin-2004} estimate the contribution (by numbers) of LSBs to gas-rich
($M_{HI}>10^8\,M_{\odot}$)
galaxies to be 50-60 per cent (using two
different methods for the estimate). Following \cite{minchin-2004},
LSBs contribute approximately 30$\pm$10\,\% to the neutral hydrogen density.\\
However, the classification of galaxies into high and low surface brightness
galaxies (HSBs \& LSBs) is not a strict separation. 
In this paper, we will classify all galaxies with a
central, inclination corrected blue surface brightness of ${\rm\mu_{B,corr}>22.5\,mag\,arcsec^{-2}}$
(i.e., $\ge 3\sigma\,$ fainter than the Freeman value) as low
surface brightness galaxies.\\
As stated before, optical surveys show a severe selection effect
regarding the detection of LSB galaxies. As this optical selection bias
does not apply to blind HI surveys, they are
considered a
good alternative in the search for LSB galaxies. Moreover, they allow a different
probe of the galaxy population. However, HI surveys
will also show some kinds of selection effects. 
Although LSBs are often regarded as gas-rich, having total HI
masses which are comparable to HSB galaxies
\citep[e.g.,][]{deblok-1996b, burkholder-2001, oneil-2004a}, blind HI surveys will miss
gas poor galaxies.
As there is a trend to low HI column densities for LSB galaxies
\citep[][]{deblok-1996b, minchin-2003}, very deep HI surveys are needed
to be able to easily detect LSBs. For column densities below
$\rm{n_{HI}\leq10^{19.7}cm^{-2}}$, ionisation of the neutral hydrogen
may become important \citep[][]{sprayberry-1998} and the amount of HI, and
therefore also the detectability in HI, will
decrease rapidly. And HI surveys have a limit in bandwidth, and thus in radial
velocity range.
Thus, it is clear that both HI and optical surveys each in their own way are biased to some extent
against the detection of LSB galaxies and that both kind of
surveys will result in a different sampling of the LSB population\\
A blind optical follow-up observation of a region of the sky which was
initially observed in the 21 cm line allows one to make a direct comparison
between an optically and an HI selected sample.
We have therefore made a blind optical follow-up observation of a part of the
region covered by the blind Arecibo HI Strip
Survey \citep[AHISS,][]{zwaan-1997}.\\
The paper is organised as follows.
Section 2 presents our data and the data reduction. In section 3, the
object detection, the catalogue handling and the selection of the sample
are described. Additionally, this section deals with the galaxy profile
fitting which was done using Galfit \citep[][]{peng-2002}.
We estimate the detection efficiency and the bias due to our
selection criteria in section 4, using intensive studies on artificial galaxies.
Section 5 deals with the comparisons to other optical surveys and to
the AHISS itself. Finally, we present our conclusions in section 6.


\section{Observations and data reduction}

\subsection{Observations}
The data were obtained at the Calar Alto Observatory in Spain. In October 1999, the 
1.23m telescope with the focal reducer WWFPP and the blue optimised
2048\,x\,2048 CCD Site\#18b was used for the \emph{B}-band data.
The FoV was 25\arcm\,x\,21\arcm\, and the image scale 1.147
arcsec/pixel.
In nine nights of good conditions we observed 15.5 deg${\rm^{2}}$ in Johnson
\emph{B} with an exposure time of 900\,s per FoV. The blooming of the used
CCD reduces the effective area to roughly 14.3 deg${\rm^{2}}$.
The  \emph{R}-band data were obtained in August 2002 at the 2.2m telescope at 
Calar Alto using CAFOS with a 2048\,x\,2048 pixel CCD
and a Roeser \emph{R2} filter. 
As a consequence of the long readout time of the CCD camera, part of the 
observations was done using a 2\,x\,2 binning to save observing
time. The image scales are 0.53  
and 1.06 arcsec/pixel and the exposure times 300\,s and
120\,s respectively. The survey area in Roeser \emph{R2} is 7.25 deg${\rm^{2}}$.
The area covered by our survey can be roughly described as two long strips out of the galactic plane at a fixed 
declination 
of 14$^{\circ}$12\arcm. The right ascension coverage of the strips is
due to observational constraints (e.g., avoiding the galactic plane,
effective use of observing time).
The exposed regions are shown in Table~\ref{Strips}.
For both passbands, standard stars from the catalogue of \cite{landolt-1992} were observed each night. 

\begin{table}[htbp]
\begin{minipagetbl}{7cm}
\caption{Area covered from our survey. The declination for all
      strips is 14$^{\circ}$12\arcm.}\label{Strips} \vspace{.1in}
\begin{tabular}{ccc} \hline \hline
            Filter &  \rm ${\alpha_{start}}$ & \rm ${\alpha_{end}}$ \\
            & J2000 & J2000 \\
            \hline
            Johnson\, \emph{B} & 21:29:00 & 22:24:00 \\
            Johnson\, \emph{B} & 22:55:15 & 00:51:45 \\
            Roeser\, \emph{R2\footnote{http://www.caha.es/CAHA/Instruments/filterlist.html}} & 21:29:30 & 22:17:00 \\
            Roeser\, \emph{R2$^{a}$} & 22:57:00 & 00:11:45 \\
            Roeser\, \emph{R2$^{a}$} & 00:12:00 & 00:29:00 \\
            \hline
\end{tabular}
\end{minipagetbl}
\end{table}
\subsection{Data reduction}
We used a data reduction pipeline for 
mosaic CCD wide field imaging data \citep[][]{erben-2005}.
The \emph{R}-band data were completely reduced by this pipeline,
including astrometric and relative photometric calibration and
mosaicing. 
For the \emph{B}-band data, only the standard reduction steps were
done using the pipeline. The astrometric calibration was done with a FORTRAN based programme of the ``Bonner Astrometrie
Programme'' \citep[BAP,][]{geffert-1997} and the relative photometry
by using \emph{SExtractor} \citep[][]{bertin-1996}.\\
An absolute photometric calibration was done after the data reduction
of the standard fields using the \emph{photcal} package of
IRAF\footnote{IRAF is the Image Reduction and Analysis Facility. IRAF is
written and supported by the IRAF programming group at the National Optical
Astronomy Observatories (NOAO) in Tucson, Arizona.}.\\
The mean error of the total photometric zeropoint -- meaning both relative and
absolute photometric calibration -- is about 0.16~mag for the images in the \emph{R} filter and 0.13 mag for 
the ones in the \emph{B} filter. This high photometric error
is mostly a result of the Gaussian error propagation in the calibration of the
relative photometry. As only a few
nights were photometric, absolute photometric calibration could be
done for a couple of exposures only. From these images, which were
roughly positioned in the middle of our strips, the relative
offsets of the photometric zeropoint were then calculated to the edges of the strips.

\section{Analysis}
In the following, some of the technical details of the analysis
concerning object detection, profile fitting of
the galaxies, and the subsequent selection criteria adopted for our
object catalogues are given.
\subsection{Detection and selections of the objects}
Our object detection was based on the \emph{B}-band and was done using
\emph{SExtractor}, using object selection criteria that were applied
uniformly over the whole data set.
The use of artificial galaxies, as shown in section 4 or in \cite{flint-2001}, yields the possibility for 
an estimation of the detection efficiency.
Another advantage of a programme like \emph{SExtractor} is that it is
not limited to the detection of objects, but is able to calculate many
parameters on the fly, which makes the subsequent analysis of the objects much easier.
We convolved the images using a tophat filter with a FWHM of 2\arcs\, before the
object detection and set \emph{SExtractor} to detect only objects with a
minimum area corresponding to five pixels in the \emph{B}-band data above a threshold of three sigma. To
clean the catalogues of ``bad'' objects, we removed all deblended and
saturated detections, i.e., those with a
flag value of more than two (see the \emph{SExtractor} \mbox{manual}). To reduce the computation time for subsequent analysis, we
made a star-galaxy separation via the \emph{SExtractor} keyword CLASS\_STAR
and rejected all detections with CLASS\_STAR $>$ 0.1 (where 0 $\hat{=}$
galaxy, 1 $\hat{=}$ star).

This leaves us with a total amount of 13\,388 objects, whereof 6\,391 objects
were detected in both filters.
\subsection{Galaxy-fitting}
These 13\,388 galaxy candidates were fed into the galaxy-fitting routine
Galfit \citep[][]{peng-2002}, using it with a batch mode written by the
GEMS\footnote{http://www.mpia.de/GEMS/gems.htm}-group and slightly modified by
us.
As LSBs mostly lack strong bulges \citep[see e.g.,][]{deblok-1995,beijersbergen-1999}, a decomposition of the galaxies into bulge and
disc is not essential. Hence, we set Galfit to fit each galaxy with an exponential profile.
For a few objects, it was necessary to create mask-images to exclude regions
affected by, e.g., CCD defects.\\
\begin{figure}
\centering
\includegraphics[width=28mm]{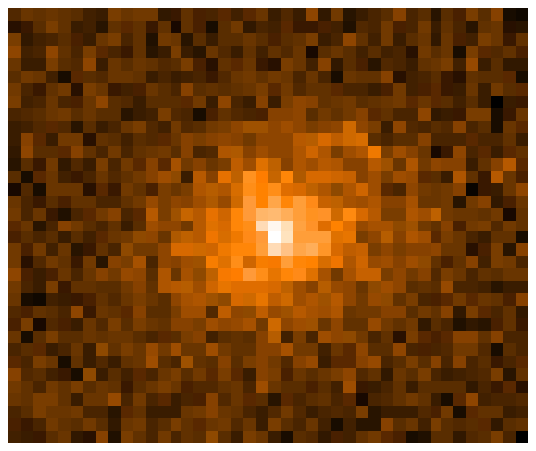}
\includegraphics[width=28mm]{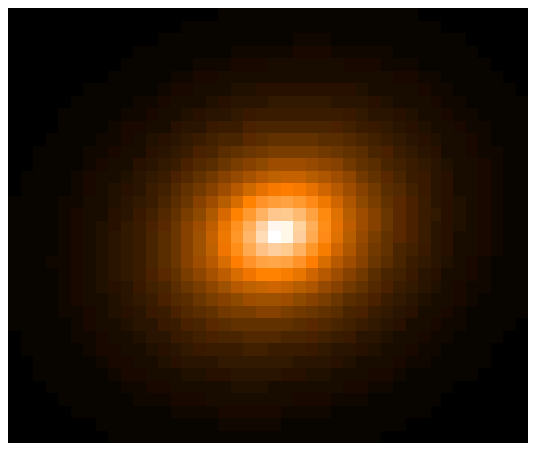}
\includegraphics[width=28mm]{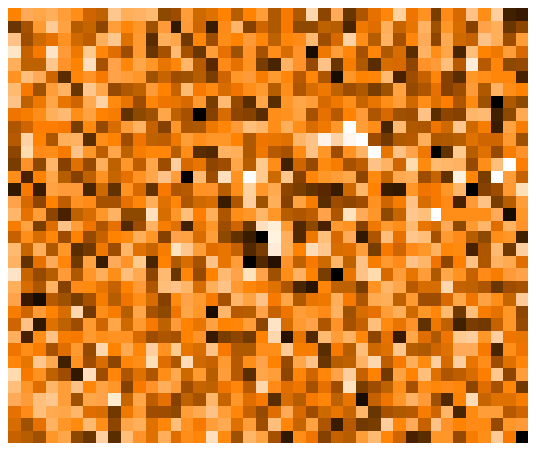}
\caption{Thumbnail triplet of one of our galaxies. Original galaxy (left),
model fit (middle) and residual image (right).}
\label{thumbnails}
\end{figure}
Galfit calculates the total magnitude, scale
length, major axis position angle and axis ratio for each galaxy and generates a triplet
of thumbnails - of the original object, the model fit and a residual
image (see Fig.~\ref{thumbnails}).
After the profile fitting, we calculated the central surface
brightness, $\rm{\mu_{B,0}}$, in units of \magarcsec\, using equation~\ref{mue0}
\citep[derived from][]{peng-2002}, where $m_B$ is the apparent, isophotal magnitude in the
\emph{B}-band, $\alpha$ is the disc scale length in pixel, q is the axis ratio and A
is the area of one pixel.
\begin{equation}
\mu_{B,0}=m_B+2.5log_{_{10}}(2\pi\alpha^{2}qA)
\label{mue0}
\end{equation}
The inclination angle, i, was obtained via equation~\ref{i} and the inclination corrected 
central surface brightness, $\rm{\mu_{B,corr}}$, was calculated using equation~\ref{mue_corr}
\citep[][]{oneil-1997a}.\\
\begin{equation}
i=cos^{-1}(q)
\label{i}
\end{equation}

\begin{equation}
\mu_{B,corr}=\mu_{B,0}-2.5\cdot log_{_{10}}(cos\,i)   
\label{mue_corr}
\end{equation}
The diameters of our objects were estimated using an analytical approach as shown in 
equation~\ref{radius}, where ${\mu_{B,lim}}$ is the limiting \emph{B}-band surface brightness
derived by \emph{SExtractor}, and ${D_{lim}}$ is the diameter in arcsec at
${\mu_{B,lim}}$. 
The use of a diameter at the faintest detectable surface brightness
  for a whole survey is
only advisable if $\mu_{B,lim}$ does not differ (very) much between the
specific exposures. Otherwise, it would lead
to very different selection criteria throughout the survey.
Our $\mu_{B,lim}$ depends on the weather conditions and the air mass. The mean value and the standard
deviation are $\mu_{B,lim}=25.2\pm0.31\,$\magarcsec. Thus, our
${D_{lim}}$ is pretty comparable to the $D_{25}$, the blue isophotal surface
brightness at a level of 25\,\magarcsec\, which is used in many
other surveys.
Note that our diameters are not estimated by eye but are based on our
  automated search and fitting algorithms. 
\begin{equation}
D_{lim}=\frac{2\,\alpha}{1.086}\cdot (\mu_{B,lim}-\mu_{B,0})
\label{radius}
\end{equation}

\subsection{Removal of the higher redshifted galaxies}\label{removal-high-z}
We want to limit our search to the local ($z\le 0.1$) universe in order to reduce the influence
of the Tolman $(1+z)^{-4}$ dimming \citep{phillipps-1990} which
shifts the central surface brightness of high redshifted galaxies to the
surface brightness region occupied by local LSBs. 
Without redshifts for all galaxies, one needs secondary methods to
reject high-z objects.
As higher redshifted galaxies usually have smaller apparent angular
scales than nearby galaxies, we used a maximum diameter rejection. 
To define a limiting diameter which rejects most of the high-z HSBs,
one needs to know the size distribution of the HSB galaxy population.
For this calibration, we used the $\rm{D_{25}}$ of the Nearby Galaxies Catalog (hereafter
NBGC) of \cite{tully-1988}, for which we assume that its size
distribution is re\-pre\-sen\-ta\-ti\-ve (at least at the upper end). 
The NBGC contains
2\,367 galaxies up to a heliocentric velocity of $cz<3\,000\,km/s$
(i.e., 40\,Mpc using $H_0=75\,km\,s^{-1}\,Mpc^{-1}$).
622 galaxies with $cz<1\,000\,km/s$ were excluded to avoid
uncertainties due to deviances from the Hubble flow.\\
\begin{table}[htbp]
\begin{minipagetbl}{7cm}
\caption{Contamination (in \%) with higher redshifted HSBs for different
limiting diameter at several survey limits.}\label{tully} \vspace{.1in}
\begin{tabular}{l|cccc} \hline \hline
            \Lower{$z_{lim}$} &
            \multicolumn{4}{c}{$D_{lim}$}\\
& 10\arcsec & 14\arcsec & 18\arcsec & 22\arcsec\\
            0.05 & 89 & 75 & 58 & 42\\
            0.06 & 82 & 63 & 44  & 27\\
            0.07 & 75 & 52 & 31 & 16\\
            0.08 & 66  & 40 & 21 & 10 \\
            0.09 & 58 & 31 & 13 & 6 \\
            0.1 & 50 & 22 & 9  & 4 \\
            \hline
\end{tabular}
\end{minipagetbl}
\end{table}\\
We converted the absolute size distribution in kpc to an
apparent one (in arcsec), by artificially shifting all galaxies to a specific
redshift $z_{lim}$. The fraction of galaxies with an apparent diameter
exceeding the limiting diameter $D_{lim}$ then gives the possible
contamination of galaxies with $z>z_{lim}$. Table~\ref{tully} shows
this contamination for various values of $z_{lim}$ and $D_{lim}$.
It is obvious that a too small $D_{lim}$ increases the contamination
with high-z HSBs, whereas a too large $D_{lim}$ would severely reject
LSBs as their $D_{25}$ is in general approximately two scale
lengths smaller than that from
HSBs - assuming the same disc scale length \citep[][]{mcgaugh-1994b} and distance distribution for HSBs
and LSBs. Using $D_{lim}=18\arcs$ seems the best compromise between the
two extremes. At an arbitrary survey limit of $z=0.1$, this $D_{lim}$
results in a contamination with high-z HSBs of about 10\,\%.
Nevertheless, the use of a diameter limit acts as a makeshift solution. To
really avoid the rejection of LSBs, one needs redshifts for the
complete sample.\\
After the removal of the objects with a diameter smaller than
18\arcs, we verified all our remaining objects by a visual inspection. All
unsatisfactorily fitted objects and CCD artifacts were 
rejected. The cleaned catalogue consists of 306 galaxies, of which 174
were also fitted in the \emph{R}-Band, and of which 148 were
previously uncatalogued. An excerpt of the catalogue is given in
table~\ref{catalogue}. The full catalogue is available in electronic
form at the CDS via anonymous ftp to cdsarc.u-strasbg.fr.
\begin{table*}[htbp]
\setlength{\tabcolsep}{0.05in}
\scriptsize
\begin{minipagetbl}{18cm}
\caption{An excerpt of the object catalogue. The full catalogue is
electronically available in the online material.}\label{catalogue} 
\begin{tabular}{ccccccccccccccc} \hline
     ID &  Other ID & \rm ${\alpha_{J2000}}$ & \rm ${\delta_{J2000}}$
          & $m_B$ & $\mu_{B,0}$ & $\mu_{B,corr}$ & $i_B$ & $\alpha$ & b/a &
          $D_{lim}$ & $m_R$ & $\mu_{R,0}$ & $\mu_{R,corr}$ & $i_R$\\
(1) & (2) & (3) & (4) & (5) & (6) & (7) & (8) & (9) & (10) & (11) & (12) & (13) & (14) & (15) \\
            \hline
TBHD\_J220915+1421.6 & [ZBS97] A09 & 22:09:15 & +14:21:38 & 14.0 & 21.38 & 22.42
            & 68 & 19.61 & 0.38 & 136.6 & 13.5 & 20.55 & 21.67 & 69 \\
TBHD\_J222047+1414.0 & [ZBS97] A13 & 22:20:47 & +14:14:05 & 15.4 & 22.27 & 22.69
& 47 & 11.60 & 0.68 & 75.0 & - & - & - & - \\
TBHD\_J230556+1421.4 & UGC 12354 & 23:05:56 & +14:21:27 & 14.3 & 20.84 & 21.84
& 66 & 12.74 & 0.40 & 100.0 & 14.2 & 20.25 & 21.21 & 66 \\
TBHD\_J225833+1410.4 & -- & 22:58:33 & +14:10:24 & 19.2 & 22.62 & 24.16 & 76 & 3.96
& 0.24 & 19.2 & 17.9 & 20.88 & 22.41 & 76 \\
TBHD\_J214854+1421.0 & -- & 21:48:54 & +14:21:20 & 17.1 & 23.43 & 23.62 & 33 & 8.18
& 0.84 & 22.1 & - & - & - & - \\
TBHD\_J213119+1407.7 & -- & 21:31:19 & +14:07:43 & 18.7 & 23.04 & 24.05 & 67 & 4.61
& 0.39 & 22.8 & - & - & - & - \\
            \hline
\end{tabular}
\end{minipagetbl}
{\sc Note:} (1) IAU conformable identifier; (2) prior
identification according to the NED (If blank, the galaxy was
previously uncatalogued); (3) right ascension (J2000) in hours,
minutes, seconds; (4) declination (J2000) in degree, minutes, seconds;
(5) apparent, isophotal magnitude in the
\emph{B}-band; (6) \emph{B}-band central surface brightness in
\magarcsec; (7) inclination corrected central \emph{B}-band surface
brightness in \magarcsec; (8) inclination angle from \emph{B}-band
data in degree; (9) \emph{B}-band scale length in arcsec; (10) axis
ratio derived from the \emph{B}-band data; (11) object diameter
in arcsec in the \emph{B}-band data at the limiting surface
brightness; (12) apparent, isophotal magnitude in the \emph{R}-band; (13)
\emph{R}-band central surface brightness in \magarcsec; (14)
inclination corrected central \emph{R}-band surface brightness in
\magarcsec; (15) inclination angle from \emph{R}-band data in degree.
For the objects of which we have no \emph{R}-band data , Cols. 12-15
are left blank.
\end{table*}

\section{The detection efficiency tested using artificial objects}
We tested our detection efficiency and the effect of our selection
criteria like CLASS\_STAR $\leq$ 0.1, FLAGS $<$ 3, $D_{lim}\geq\,18\arcs$, and
masking of blooming regions with large samples of artificial galaxies. We created these
objects using the \emph{gallist} task in IRAF and added about
85\,000 galaxies of varying pa\-ra\-me\-ters in our \emph{B}-band images using the IRAF
task \emph{mkobject}. The galaxies cover a large parameter space in
total magnitude ($12\leq m_B\leq 22\,mag$) and central,
for inclination uncorrected, surface brightness
($18.5\leq \mu_{B,0} \leq 26\,$\magarcsec). We ensured that each bin
(the bin size is 0.25 mag and \magarcsec, respectively) in
both parameters contains at least 10-20 objects, and most bins (especially
in regions where the detection efficiency drops) contain more than
50 galaxies (the mean value is 90). The input parameters from  \emph{gallist} are
\emph{magnitude}, \emph{eradius}, \emph{position angle} and \emph{axis ratio}. The \emph{eradius} is
related to the scale length $\rm{\alpha}$ by equation~\ref{eradius}. 
\begin{equation}
eradius=2\alpha\,log_{10}(2)
\label{eradius}
\end{equation}
The object detection was done by
\emph{SExtractor} using the same parameters as on the original
science frames. The original objects were then excluded from the catalogues so
that the catalogues contain only the new, artificial galaxies.\\
After the removal of flagged objects (either with the blooming flag or with FLAGS
$\geq$ 3), the remaining objects were fed into Galfit (again with the 
same parameters as in the survey) and the size of each object was calculated. Due to
the huge numbers of galaxies, we refrained from checking each object
visually and simply rejected all objects which could not be
fitted by Galfit without interaction. For the
remaining galaxies, we adopted the same diameter criterion as for our
original data and rejected all galaxies smaller than 18\arcs\, in
diameter.
\begin{figure}
\centering
\hspace{-0.75cm}
\includegraphics[width=6.5cm, angle=-90]{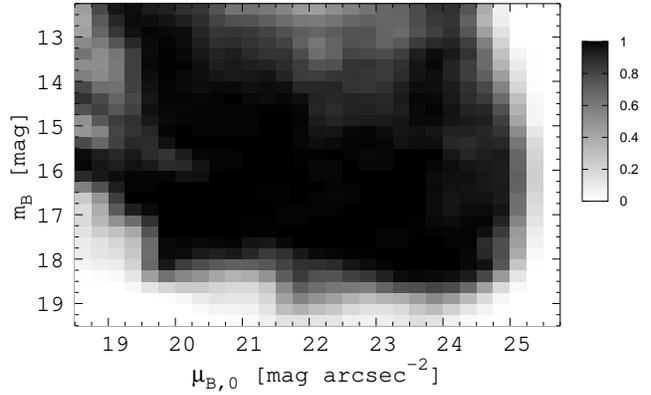}
\caption{Surface plot of the detection efficiency in respect of
magnitude and surface brightness. The dip in the efficiency for bright
objects is mainly due to the expansion of these objects into areas
affected with blooming which led to their rejection.}
\label{det-eff1}
\end{figure}
\subsection{The detection efficiency for our data}
The surface plot in Fig.~\ref{det-eff1} shows the fraction of the
re\-co\-ve\-red objects in respect of the input objects in the same
magnitude/surface brightness bin. The bin size is 0.25 mag and
\magarcsec\, respectively. 
The detection efficiency exceeds the 90\,\% level for a large part of
the parameter space ($\rm{15.5\la m_B\la 18}$ mag \& $\rm{20.5\la \mu_{B,0}\la
24}$ \magarcsec). Nevertheless, our sample of real galaxies contains only few
objects with $\rm{\mu_{B,0} \la 23\,mag\,arcsec^{-2}}$, as these objects need quite large scale length to reach a $\rm{D_{lim}>18\arcs}$.
Objects with $\rm{\mu_{B,0} \la 19.5\,mag\,arcsec^{-2}}$ look rather
star-like and were partly rejected due to ${\rm CLASS\_STAR\ge 0.1}$.
Also rejected were the bright objects with
$\rm{20.5\le\mu_{B,0}\le23.5\,}$\magarcsec\, which, due to their large sizes, have a
higher probability of being located in an area affected by blooming.\\
The drop-off towards faint surface brightnesses is quite sharp and
rather independent from the total magnitude. In contrast, the decline in magnitude
direction is more dependent on the surface brightness. The
point at which the decline occurs moves slightly towards fainter magnitudes if
one goes to fainter surface brightnesses, as the scale length for
these artificial galaxies is larger and they can achieve a
diameter larger than 18\arcs\, before disappearing into the
noise.
\subsection{The detection efficiency of the real galaxies}
\begin{figure}
\centering
\hspace{-0.75cm}
\includegraphics[width=6.5cm, angle=-90]{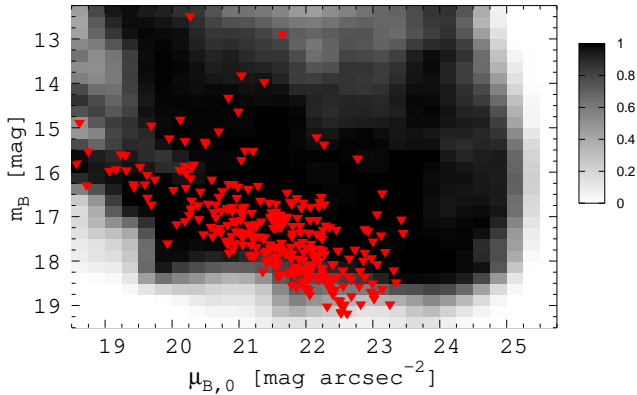}    
\caption{The same surface plot as in Fig.~\ref{det-eff1}, but 
we additionally inserted the detected objects into this plot
(red triangles). As one can see, the real galaxies do not cover the whole area
which is accessible with our search methods, and most objects lay in regions
with high detection efficiency.}
\label{det-eff2}
\end{figure}
Figure~\ref{det-eff2} shows the same as Fig.~\ref{det-eff1}, with the addition
of real objects from our sample (red triangles).
As both data sets underwent the same analysis and selection, no
systematic difference should arise between them.
The real galaxies are mostly located in a well defined region of the
diagram, even though the parameter space covered by the artificial galaxies is quite large. 
This is due to the fact that the parameter range of our artificial galaxies
is purely theoretical, and, e.g., galaxies with
a scale length of 50\arcsec\, should be rather rare.
Nevertheless, most of the found objects are located
in regions with a high detection efficiency, where our
survey is expected to be quite complete.
For faint objects with 
$\rm{m_B>18.5\,mag}$, the detection efficiency drops below 50\,\% and thus
our galaxy counts for these objects give only lower limits, which could be
more than a factor of 2 too low.
Analogous to the trends in \cite{cross-2002} and \cite{driver-2005},
we can see a general trend towards fainter central surface brightnesses if one
goes to fainter total magnitudes.

\section{Results}
\subsection{The number density}
\begin{figure}
\centering
\includegraphics[width=8.5cm, angle=0]{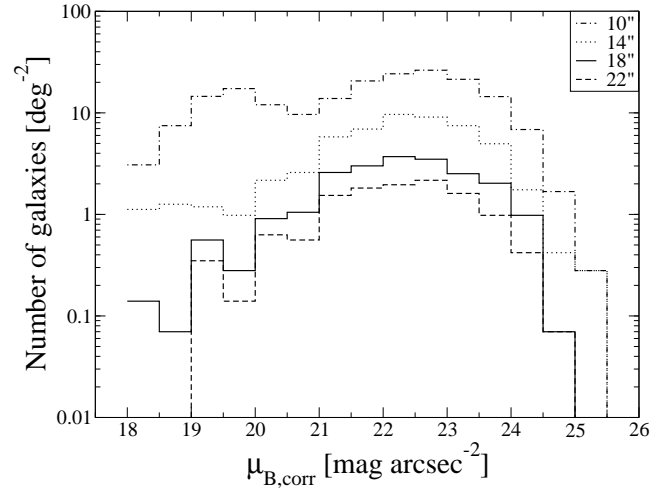}
\caption{The number of galaxies per deg$^{2}$ per
inclination corrected central surface 
brightness bin (bin size 0.5 \magarcsec) for samples with different
limiting diameters is shown on a logarithmic scale. The limiting
diameter has a strong influence on the number density.}
\label{anzahldichte}
\end{figure}
Figure~\ref{anzahldichte} shows the distribution of the inclination
corrected central surface brightness ($\mu_{B,corr}$) for our final
sample ($D_{lim}>18\arcs$) and for other limiting diameter. 
A smooth transition can be seen between galaxies of high and low surface
brightness, at $\rm{\mu_{B,corr}=22.5\,}$\magarcsec. For smaller $D_{lim}$,
both the number density and the
contamination with high redshifted galaxies are higher. Setting an arbitrary survey
limit at $z=0.1$ and using the assumptions based on the NBGC
as discussed in Section~\ref{removal-high-z}, a
$D_{lim}$ of 10\arcs\, leads to a 50\,\% contamination of galaxies with $z\la
0.1$, whereas at $D_{lim}=18\arcs$ this 
contamination is reduced to only 9\,\% (see Table~\ref{tully}).\\
The objects in the brightest bins ($\rm{\mu_{B,corr}<20\,mag\,arcsec^{-2}}$) are
mainly spiral galaxies with strong bulges, which are responsible for their
high central surface brightnesses, since we did not make any
decomposition but fitted a single profile to our galaxies.\\ 
We classified 130 galaxies (42\,\%) as LSBs.
Even if we restrict the survey to $z<0.1$, cosmological dimming has
to be taken into account, as at $z=0.1$ a galaxy is dimmed about
0.4~\magarcsec. 
To avoid an overestimation of LSBs, we used the most extreme case that all
galaxies are located at $z=0.1$, which shifts our LSB selection criterion from
22.5~\magarcsec\, to 22.9~\magarcsec.
If we adopt this criterion, 96 (30\,\%) of our galaxies are
LSBs, which indicates that these objects contribute substantially to the local galaxy number density, since this number is clearly a lower limit.\\
Our results of 30-40\,\% are in agreement with other
surveys. \cite{minchin-2004} estimated the contribution (by numbers) of low surface
brightness galaxies to gas-rich ($M_{HI}>10^8\,M_{\odot}$)
galaxies to be 50-60\,\%, and 50\,\% of the galaxies in the sample of \citet[][]{spitzak-1998}
have a central disc surface brightness fainter than 22.5 \magarcsec\,
in the \emph{B}-band and are therefore LSBs. In the sample of
\citet[][]{driver-2005}, 26\,\% of the galaxies have an effective
surface brightness fainter than 23.6 \magarcsec. According to
\citet[][]{cross-2001}, this corresponds to a
central surface brightness of 22.5 \magarcsec.\\
\subsection{Comparison with other optical surveys}
In the following, we compare the surface brightness distribution of our survey with
that from two other optical surveys. The first one is the Texas-Survey
by \citet[][ OBC97 hereafter]{oneil-1997a} and the second one is
a very deep search for LSBs in the HDF--S \citep[][HBDP hereafter]{haberzettl-2006}.
\begin{figure}[h]
\centering
\includegraphics[width=8.5cm]{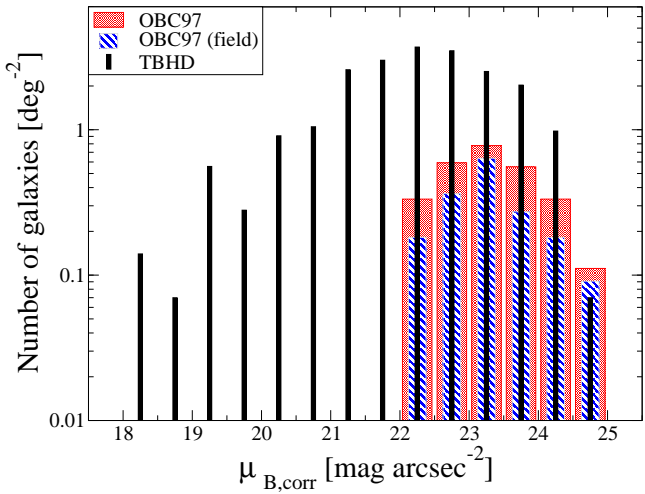}
\caption{The SB distribution for our survey (TBHD) and for OBC97 and the
field sample of OBC97. To make a comparison of the two surveys easier,
the same diameter limit was applied to all samples (see text).}
\label{oneil-vergleich}
\end{figure}
\\
\subsubsection{Texas-Survey (OBC97)}
The Texas-Survey was targeted on cluster and field
environments and covered an area of 27~deg$^2$. The original catalogue contained 127 galaxies which
were visually selected. The selection criteria of OBC97 were
$\rm{\mu_{B,0}\ge\,22\, mag\,arcsec^{-2}}$ and $R_{27}\ga
13\farcs2$. The size estimation was also done visually, and the limiting
surface brightness of the survey is
$\rm{\mu_{lim}\approx\,27\,mag\,arcsec^{-2}}$. It is not straightforward to
compare two surveys with such differences in the object detection, 
size estimation, and adopted diameter limits. To make the comparison
easier, we compared samples with almost identical diameter limits, and instead of using $R_{27}\ga
13\farcs2$ we used $D_{25}>18\arcs$, which matches our
diameter criterion. Moreover, we made an additional analysis using
only the field objects of OBC97, as galaxy clusters typically show a
different galaxy population than field galaxies
\citep[][]{binggeli-1987}.\\
The SB distribution of our survey (referred to as TBHD) and of OBC97 and the field sample of
OBC97 is shown in Fig.~\ref{oneil-vergleich}.
The decline of the OBC97 sample in the $\rm{22\le\mu_{B,corr}\le
22.5\,mag\,arcsec^{-2}}$ bin is a result of their upper SB selection
limit. Between $\rm{22.5\le\mu_{B,corr}\le 24.5\,mag\,arcsec^{-2}}$, our number
density is higher than that of
OBC97 (in average 6.5 times as high as the field sample and 3.9 times
as high as the field+cluster sample).
The differences in the number density and the survey areas are high enough to exclude cosmic
variance (our survey area: 14.3~deg$^2$, OBC97: 27~deg$^2$, of which
$\approx\,$11~deg$^2$ is field). It may be possible that the differences
originate from the diverse methods for object detection and size
estimation (which is important if using a diameter limit). If this were
true, it would show that automated methods for object detection and
size estimation increase the galaxy number density significantly - in any case, it emphasises
that galaxy catalogues created by automated methods are at least as
complete as visually selected ones.\\
The relatively low (compared
to OBC97) number density in the faintest populated bin may be a result from
our automated star-galaxy separation which becomes difficult and partially
uncertain for low S/N objects.\\
\subsubsection{HDF--S (HBDP)}
\begin{figure}[h]
\centering
\includegraphics[width=8.5cm, angle=0]{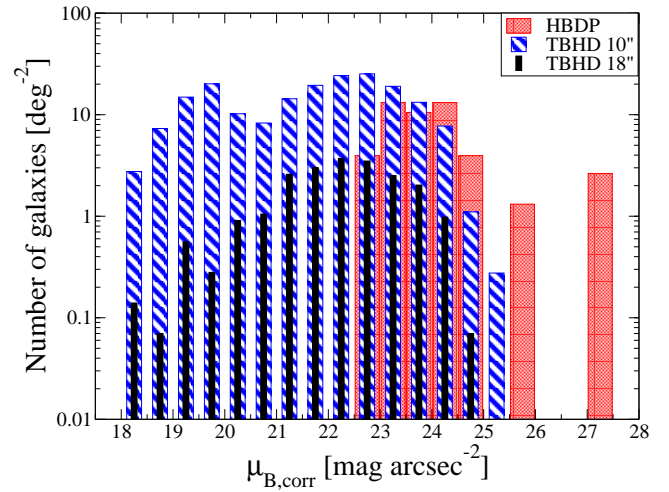}
\caption{As Figure 5; the SB distribution of HBDP and for two diameter limits
of our survey (TBHD).}
\label{haberzettl-vergleich}
\end{figure}
Figure~\ref{haberzettl-vergleich} shows the comparison between our survey and
the sample from HBDP.
The survey which HBDP used has a
limiting surface brightness of $\rm{\mu_{B_w}\approx 29\,
mag\,arcsec^{-2}}$ in the
$B_w$-band\footnote{http://www.noao.edu/kpno/mosaic/filters/filters.html},
which is slightly broader than Johnson-\emph{B}. HBDP estimate the
$<\mu_{B_w}-\mu_B>$ correction term for their SB as $0.41\pm 0.28$
\magarcsec\,and adopted this
offset to their objects without data in Johnson-\emph{B}. For
Fig.~\ref{haberzettl-vergleich} we used the inclination
corrected central surface brightnesses in Johnson-\emph{B}.
The object detection of HBDP was done by \emph{SExtractor} and they included
only galaxies in their catalogue with $\mu_{B_w}>22\,mag\,arcsec^{-2}$
and a visually estimated diameter at the limiting isophote of more
than 10\farcs8. Additionally to the \emph{SExtractor} search, HBDP used a median filter
method adopted from \citet[][]{armandroff-1998}. HBDP removed all
objects found by \emph{SExtractor} from the original data and filtered
the source-free image with a median filter with a kernel size of 25
pixel (10\farcs8). With this method, they detected the three extreme LSB
objects
which populate the bins with $\rm{\mu_{B,corr}>25.5\,mag\,arcsec^{-2}}$.\\
As a comparison of a $D_{29}>10\farcs8$ sample with a $D_{25}>18\arcs$
sample is not straightforward, 
we added a third sample with $D_{25}>10\arcs$ to
Fig.~\ref{haberzettl-vergleich}, although even the latter limit is clearly different from $D_{29}>10\arcs$. 
Although HBDP measure their diameters at a four magnitudes fainter SB
level, the number density of their sample for
surface brightnesses between $\rm{23\le \mu_{B,corr} \le
24.5\,mag\,arcsec^{-2}}$ is only as high as that of our $D_{25}>10\arcs$
sample.
If HBDP had also used $D_{25}>10\arcs$ for their selection, their
number density would have dropped by a factor of more than 2.5 and be significantly below ours.\\
Cosmic variance and low number statistics may play a
role here as the area of HBDP and their number of galaxies are quite small
(37 galaxies in 0.76~deg$^2$). Another possible reason may again
be the different methods for the size estimation. It is
possible that the eye underestimates galaxy sizes, which, when using a diameter
limit would lead to a lower number density.\\
This hypothesis is supported by first tests we made using a \emph{SExtractor}-based
search and a galaxy fitting using Galfit on the HDF--S image of
HBDP. Adopting their selection criteria
($\rm{\mu_{B_w}>22\,mag\,arcsec^{-2}, D_{lim}>10\farcs8}$) and estimating
the size analytically instead of visually increased the number of
galaxies matching these criteria significantly (up to a factor of $\approx$ 5).
\begin{table*}[htbp]
\setlength{\tabcolsep}{0.05in}
\scriptsize
\begin{minipagetbl}{18cm}
\caption{Optical and HI parameter for the galaxies found in both our
survey and in the AHISS.}\label{tbhd-ahiss} 
\begin{tabular}{rcccccccccc} \hline
     ID & \rm ${\alpha_{J2000}}$ & \rm ${\delta_{J2000}}$
          & $m_B$ & $\mu_{B,0}$ & $\mu_{B,corr}$ &  $i_B$ & $D_{lim}$
          & $M_B^{b}$ & log $M_{HI}$ & $M_{HI}/L_B^{b,i}$\\
(1) & (2) & (3) & (4) & (5) & (6) & (7) & (8) & (9)$^*$ & (10)$^*$ & (11)$^*$\\
            \hline
ZBS97\, A03 &   21:49:27 & +14:14:00 & 15.7 & 22.77 & 23.24 & 50 &
          50.61 & -15.8 & 9.1 & 3.84\\
ZBS97\, A04-1 & 21:59:05 & +14:14:19 & 17.4 & 23.01 & 23.28 & 39 &
          27.61 & -13.6 & 8.0 & 1.99\\
$\rm{[ZBS97\, A04}$-$\rm{2]}$\footnote{Galaxy affected by blooming. Given optical values
          should be treated with care.} & 21:58:35 & +14:06:53 & 14.9 & 22.03 & 22.25 & 35 &
          75.07 & -16.3 & 9.2 & 2.81\\
ZBS97\, A09 &   22:09:15 & +14:21:38 & 14.0 & 21.38 & 22.42 & 68 &
          136.56 & -16.7 & 8.8 & 0.64\\
ZBS97\, A10 &   22:58:10 & +14:18:31 & 12.9 & 21.65 & 21.70 & 18 &
          153.47 & -18.2 & 8.9 & 0.27\\
ZBS97\, A11 &   23:01:18 & +14:20:22 & 13.8 & 21.04 & 22.59 & 76 &
          173.77 & -17.5 & 9.5 & 1.54\\
$\rm{[ZBS97\, A12]}$\footnote{Rejection due to CLASS\_STAR $>$ 0.1} &   23:26:09 & +14:15:46 & 19.6 & 23.23 & 23.94 & 58 &
          8.71 & -12.7 & 7.9 & 4.21\\
ZBS97\, A13 &   22:20:47 & +14:14:05 & 15.4 & 22.27 & 22.69 & 47 &
          75.01 & -17.5 & 9.3 & 1.31\\
ZBS97\, A14 &   23:36:37 & +14:09:25 & 15.2 & 22.16 & 22.90 & 60 &
          81.10 & -17.8 & 9.7 & 1.94\\
ZBS97\, A15 &   00:11:08 & +14:14:22 & 16.8 & 22.19 & 22.99 & 61 &
          35.98 & -12.2 & 7.3 & 1.57\\
ZBS97\, A17 &   00:20:09 & +14:17:28 & 16.4 & 21.14 & 21.84 & 58 &
          34.20 & -16.7 & 9.2 & 2.02\\
$\rm{[ZBS97\, A18]}$$^a$ &   00:24:30 & +14:16:15 & 16.6 & 21.71 & 23.11 & 74 &
          51.40 & -16.9  & 8.5 & 0.26\\
ZBS97\, A19 &   00:28:03 & +14:18:07 & 17.2 & 21.91 & 22.08 & 31 &
          23.36 & -15.4 & 8.9 & 3.48\\
$\rm{[ZBS97\, A21]}$\footnote{Image of this galaxy has poor S/N
          ($\mu_{lim}\approx24.15$ only), $D_{lim}$ of this galaxy $<$
          18\arcs} &   00:44:24 & +14:17:55 & 17.5 & 20.95 & 21.26 & 41 &
          13.54 & -15.1 & 8.3 & 0.83\\
            \hline
\end{tabular}
\end{minipagetbl}
{\sc Note:} (1) identifier according to ZBS97; (2) right
ascension (J2000) in hours, minutes, seconds; (3) declination (J2000)
in degree, minutes, seconds; (4) apparent, isophotal magnitude in the
\emph{B}-band; (5) \emph{B}-band central surface brightness in
\magarcsec; (6) inclination corrected central \emph{B}-band surface
brightness in \magarcsec; (7) the inclination angle obtained from the
\emph{B}-band data in degree; (8) object diameter
in arcsec in the \emph{B}-band data at the limiting surface
brightness; (9) Absolute \emph{B}-band magnitude corrected for
Galactic extinction, not corrected for inclination effects; (10)
Logarithm of HI mass in solar masses; (11) Ratio of HI mass to \emph{B}
luminosity in solar units. The luminosity is corrected for inclination
effects using the method of \cite{tully-1998}.\\ 
$^*$   Columns 9-11 and their description are taken
from \cite{zwaan-phd}. 
\end{table*}
\subsection{Comparison with the AHISS}
\citet[][ZBS97 hereafter]{zwaan-1997} used the AHISS to search for 
extragalactic sources and found 66 galaxies up to a velocity of 
${\rm cz=7400\, km\,s^{-1}}$ in an area of 65 deg${^{2}}$, of which 
51 were detected by the main beam
of the Arecibo telescope, which covered an area of 15
deg$^{2}$. The telescope sidelobes are considerably less sensitive than the main beam, and their sensitivity is uncertain due to temperature dependencies and
asymmetries \citep[][]{schneider-1998}. Thus, we will henceforth restrict
our comparison to the main beam of the Arecibo telescope, unless
otherwise stated.
The AHISS is
until now the most sensitive blind HI survey in terms of HI mass
and flux limits. The average noise level of the AHISS is 0.75~mJy for
a velocity resolution of 16~km\,s$^{-1}$ and the HI mass limit at the full
survey depth is $\rm{1.5\,\times\,10^8\,h^{-2}\,M_\odot}$ (ZBS97) - assuming a
profile width of $\rm{85\,km/s}$ and a sigma level of 5.\\
For the direct cross-check of the detections in the blind AHISS with those in
our survey, we included also the sidelobe detections.
The AHISS contains 15 objects in the area which is covered by
our survey. 
All 15 sources were found in our images by our automated routines,
but the five whose names appear in square brackets in Table~\ref{tbhd-ahiss} were rejected during the data processing.
The optical and HI properties of 14 of these galaxies are given in
Table~\ref{tbhd-ahiss}.
Of the five rejected galaxies, three are affected by blooming
(A04-2, A18, A20) - the latter so badly that it was omitted from
Table~\ref{tbhd-ahiss}, a fourth (A21) has a $D_{lim}$ of only 13\farcs5, and a fifth (A12) which is quite faint and small is located in an area with low
S/N and was
given a CLASS\_STAR parameter of 0.31 by \emph{SExtractor}.\\
A comparison of the main beam sample of ZBS97 with our sample allows more
general comments.
The galaxy number density in the main beam of the AHISS is 3.4 per
deg$^{2}$ (51 in 15 deg$^{2}$), whereas ours is
21.4 per deg$^{2}$ (306 in 14.3 deg$^{2}$).
Thus, our number density (for HSBs + LSBs) is a factor of 6.3 higher
than that from the AHISS.
If we make this comparison for the LSB samples only, the difference in
number density decreases.
Taking the surface brightness values of the AHISS sample from \cite{zwaan-phd},
and using ${\rm \mu_{B,corr}>22.9\,mag\,arcsec^{-2}}$ to define LSBs in our
catalogue (i.e., adopting a
cosmological dimming of 0.4 \magarcsec\, for our estimated survey limit at
$z=0.1$), the difference decreases to a factor of four (number density TBHD
6.7, AHISS 1.67 galaxies per deg$^{2}$). If we neglect all
cosmological corrections and use ${\rm \mu_{B,corr}>22.5\,mag\,arcsec^{-2}}$, the
number densities differ by a factor of 5.5. 
If dimming could be applied correctly (i.e., when the redshifts for all
objects are known), it is likely that the factor will lie between 4
and 5.5.\\
It is not unexpected that the number density of our optical survey is
significantly larger than that of the AHISS, as the AHISS has a limited velocity
range. To estimate to what extent the velocity range contributes to
this differences, one would need redshifts for all (or at least for most)
of the optically selected \mbox{galaxies}.\\
\noindent
The difference in the number
densities is smaller for the LSB subsample (4-5.5) than for the full sample
(6.3).
This trend may have two reasons. Firstly, HI surveys may have a smaller bias
against the detection of LSBs as optical surveys (regardless of the
fact that all sources detected by AHISS have optical counterparts).
Secondly, our selection criteria - especially the diameter limit of
$D_{lim}\approx D_{25} >18\arcs$ - may cause the rejection of
LSBs. However, using a smaller diameter
limit will artificially increase the number of LSBs due to
cosmologically dimmed HSB galaxies.
Thus, without spectroscopic follow-up observations, diameter-limited catalogues
created from optical surveys will either be strongly biased against LSBs, or
contaminated with high redshifted HSBs.

\section{Conclusions}
We performed optical follow-up observations of a part of the AHISS, and detected optical counterparts of all HI detections. We estimated the detection efficiency of our survey using
large samples of artificial galaxies. The
comparison of our survey with two other optical surveys
indicates that an automated search algorithm and an analytical size
estimation may increase the number of galaxies per
deg$^2$ compared to catalogues based on visual estimates. We show that diameter-limited catalogues created
by automated routines are at least as complete as visually created ones.\\
Although
our number density is higher than that of the AHISS (mainly due to the
limited velocity range of AHISS), the fraction of LSBs in the AHISS is higher
than in our survey (30-42\,\% in TBHD vs. 49\,\% in AHISS).
We suggest that this is caused by our relatively large diameter
limit, which reduces the contamination of the sample with
cosmologically dimmed higher redshifted galaxies, but also rejects local LSBs.
In order to reduce this bias against the selection of LSBs in optical surveys
one needs redshifts for the whole sample.
For a given observing time, optical surveys are despite all drawbacks best
suited to detect a large number of galaxies. The
information which optical surveys provide on detected galaxies is, however, fundamentally different from that of HI surveys. The radial velocity information inherent to HI surveys allows a direct estimation of the
volume density and the rejection of high-z HSBs. However, they
will miss the gas-poor LSBs and can only probe a
limited velocity range. 
With current telescopes, only very massive galaxies
($M_{HI}>10^{10}\,M_\odot$) can be detected beyond 10\,000
km\,s$^{-2}$. For example, 
the 6$\sigma$ HI mass limit of the ALFALFA survey will be $\sim
9.5\,\times\,10^9\,M_{\odot}$ at $cz=10\,000\,km/s$ \cite[][]{giovanelli-2005}.
Thus, HI and optical surveys are complementary, as are the differences between the samples of LSB galaxies they detect.
Therefore, we especially need extremely deep surveys of both kinds to extend our - still
limited - know\-ledge about LSBs. That we have not yet reached \emph{the
end} can e.g., be seen from the very deep HBDP optical survey.

\begin{acknowledgements}
The authors want to thank Erwin de Blok, Janine van Eymeren, and Martin Zwaan for many
 fruitful discussions and comments, which helped improving
 this paper.
Moreover, we thank Michael Geffert for his aid with the astrometric
 calibration and Giuseppe
 Aronica for acting as a replacement observer of the \emph{R}-band
 data on short notice.
Finally, we thank the referee, Wim van Driel, whose detailed comments significantly
 improved this paper.
This work was supported by the German Ministry for Education and
 Science (BMBF) under  project 05~AV5PDA/3 and by the
 Deutsche Forschungsgemeinschaft (DFG) through grant BO1642/2-1.
It is based on observations collected at the Centro Astron$\rm{\acute{o}}$mico Hispano
Alem$\rm{\acute{a}}$n (CAHA) at Calar Alto.
This research made use of the NASA/IPAC Extragalactic Database
(NED) which is operated by the Jet Propulsion Laboratory, California
Institute of Technology, under contract with the National Aeronautics
and Space Administration.
\end{acknowledgements}
\bibliographystyle{aa}
\bibliography{4545bib.bib}

\begin{thebibliography}{39}
\expandafter\ifx\csname natexlab\endcsname\relax\def\natexlab#1{#1}\fi

\bibitem[{{Armandroff} {et~al.}(1998){Armandroff}, {Davies}, \&
  {Jacoby}}]{armandroff-1998}
{Armandroff}, T.~E., {Davies}, J.~E., \& {Jacoby}, G.~H. 1998, \aj, 116, 2287

\bibitem[{{Beijersbergen} {et~al.}(1999){Beijersbergen}, {de Blok}, \& {van der
  Hulst}}]{beijersbergen-1999}
{Beijersbergen}, M., {de Blok}, W.~J.~G., \& {van der Hulst}, J.~M. 1999, \aap,
  351, 903

\bibitem[{{Bertin} \& {Arnouts}(1996)}]{bertin-1996}
{Bertin}, E. \& {Arnouts}, S. 1996, \aaps, 117, 393

\bibitem[{{Binggeli} {et~al.}(1987){Binggeli}, {Tammann}, \&
  {Sandage}}]{binggeli-1987}
{Binggeli}, B., {Tammann}, G.~A., \& {Sandage}, A. 1987, \aj, 94, 251

\bibitem[{{Bothun} {et~al.}(1997){Bothun}, {Impey}, \& {McGaugh}}]{bothun-1997}
{Bothun}, G., {Impey}, C., \& {McGaugh}, S. 1997, \pasp, 109, 745

\bibitem[{{Burkholder} {et~al.}(2001){Burkholder}, {Impey}, \&
  {Sprayberry}}]{burkholder-2001}
{Burkholder}, V., {Impey}, C., \& {Sprayberry}, D. 2001, \aj, 122, 2318

\bibitem[{{Cross} \& {Driver}(2002)}]{cross-2002}
{Cross}, N. \& {Driver}, S.~P. 2002, \mnras, 329, 579

\bibitem[{{Cross} {et~al.}(2001){Cross}, {Driver}, {Couch}, {Baugh},
  {Bland-Hawthorn}, {Bridges}, {Cannon}, {Cole}, {Colless}, {Collins},
  {Dalton}, {Deeley}, {De Propris}, {Efstathiou}, {Ellis}, {Frenk},
  {Glazebrook}, {Jackson}, {Lahav}, {Lewis}, {Lumsden}, {Maddox}, {Madgwick},
  {Moody}, {Norberg}, {Peacock}, {Peterson}, {Price}, {Seaborne}, {Sutherland},
  {Tadros}, \& {Taylor}}]{cross-2001}
{Cross}, N., {Driver}, S.~P., {Couch}, W., {et~al.} 2001, \mnras, 324, 825

\bibitem[{{de Blok} {et~al.}(1996){de Blok}, {McGaugh}, \& {van der
  Hulst}}]{deblok-1996b}
{de Blok}, W.~J.~G., {McGaugh}, S.~S., \& {van der Hulst}, J.~M. 1996, \mnras,
  283, 18

\bibitem[{{de Blok} {et~al.}(1995){de Blok}, {van der Hulst}, \&
  {Bothun}}]{deblok-1995}
{de Blok}, W.~J.~G., {van der Hulst}, J.~M., \& {Bothun}, G.~D. 1995, \mnras,
  274, 235

\bibitem[{{Disney}(1976)}]{disney-1976}
{Disney}, M.~J. 1976, \nat, 263, 573

\bibitem[{{Driver} {et~al.}(2005){Driver}, {Liske}, {Cross}, {De Propris}, \&
  {Allen}}]{driver-2005}
{Driver}, S.~P., {Liske}, J., {Cross}, N.~J.~G., {De Propris}, R., \& {Allen},
  P.~D. 2005, \mnras, 360, 81

\bibitem[{{Erben} {et~al.}(2005){Erben}, {Schirmer}, {Dietrich}, {Cordes},
  {Haberzettl}, {Hetterscheidt}, {Hildebrandt}, {Schmithuesen}, {Schneider},
  {Simon}, {Deul}, {Hook}, {Kaiser}, {Radovich}, {Benoist}, {Nonino}, {Olsen},
  {Prandoni}, {Wichmann}, {Zaggia}, {Bomans}, {Dettmar}, \&
  {Miralles}}]{erben-2005}
{Erben}, T., {Schirmer}, M., {Dietrich}, J.~P., {et~al.} 2005, Astronomische
  Nachrichten, 326, 432

\bibitem[{{Flint} {et~al.}(2001){Flint}, {Metevier}, {Bolte}, \& {Mendes de
  Oliveira}}]{flint-2001}
{Flint}, K., {Metevier}, A.~J., {Bolte}, M., \& {Mendes de Oliveira}, C. 2001,
  \apjs, 134, 53

\bibitem[{{Freeman}(1970)}]{freeman-1970}
{Freeman}, K.~C. 1970, \apj, 160, 811

\bibitem[{{Geffert} {et~al.}(1997){Geffert}, {Klemola}, {Hiesgen}, \&
  {Schmoll}}]{geffert-1997}
{Geffert}, M., {Klemola}, A.~R., {Hiesgen}, M., \& {Schmoll}, J. 1997, \aaps,
  124, 157

\bibitem[{{Giovanelli} {et~al.}(2005){Giovanelli}, {Haynes}, {Kent},
  {Perillat}, {Saintonge}, {Brosch}, {Catinella}, {Hoffman}, {Stierwalt},
  {Spekkens}, {Lerner}, {Masters}, {Momjian}, {Rosenberg}, {Springob},
  {Boselli}, {Charmandaris}, {Darling}, {Davies}, {Lambas}, {Gavazzi},
  {Giovanardi}, {Hardy}, {Hunt}, {Iovino}, {Karachentsev}, {Karachentseva},
  {Koopmann}, {Marinoni}, {Minchin}, {Muller}, {Putman}, {Pantoja}, {Salzer},
  {Scodeggio}, {Skillman}, {Solanes}, {Valotto}, {van Driel}, \& {van
  Zee}}]{giovanelli-2005}
{Giovanelli}, R., {Haynes}, M.~P., {Kent}, B.~R., {et~al.} 2005, \aj, 130, 2598

\bibitem[{{Haberzettl} {et~al.}(in press.){Haberzettl}, {Bomans}, {Dettmar}, \&
  {Pohlen}}]{haberzettl-2006}
{Haberzettl}, L., {Bomans}, D.~J., {Dettmar}, R.-J., \& {Pohlen}, M. in press.,
  Low Surface Brightness Galaxies around the HDF-S: I. Object extraction and
  photometric results

\bibitem[{{Impey} \& {Bothun}(1997)}]{impey-1997-review}
{Impey}, C. \& {Bothun}, G. 1997, \araa, 35, 267

\bibitem[{{Impey} {et~al.}(1988){Impey}, {Bothun}, \& {Malin}}]{impey-1988}
{Impey}, C., {Bothun}, G., \& {Malin}, D. 1988, \apj, 330, 634

\bibitem[{{Landolt}(1992)}]{landolt-1992}
{Landolt}, A.~U. 1992, \aj, 104, 340

\bibitem[{{McGaugh}(1996)}]{mcgaugh-1996}
{McGaugh}, S.~S. 1996, \mnras, 280, 337

\bibitem[{{McGaugh} \& {Bothun}(1994)}]{mcgaugh-1994b}
{McGaugh}, S.~S. \& {Bothun}, G.~D. 1994, \aj, 107, 530

\bibitem[{{McGaugh} {et~al.}(1995){McGaugh}, {Schombert}, \&
  {Bothun}}]{mcgaugh-1995}
{McGaugh}, S.~S., {Schombert}, J.~M., \& {Bothun}, G.~D. 1995, \aj, 109, 2019

\bibitem[{{Minchin} {et~al.}(2003){Minchin}, {Disney}, {Boyce}, {de Blok},
  {Parker}, {Banks}, {Freeman}, {Garcia}, {Gibson}, {Grossi}, {Haynes},
  {Knezek}, {Lang}, {Malin}, {Price}, {Stewart}, \& {Wright}}]{minchin-2003}
{Minchin}, R.~F., {Disney}, M.~J., {Boyce}, P.~J., {et~al.} 2003, \mnras, 346,
  787

\bibitem[{{Minchin} {et~al.}(2004){Minchin}, {Disney}, {Parker}, {Boyce}, {de
  Blok}, {Banks}, {Ekers}, {Freeman}, {Garcia}, {Gibson}, {Grossi}, {Haynes},
  {Knezek}, {Lang}, {Malin}, {Price}, {Putman}, {Stewart}, \&
  {Wright}}]{minchin-2004}
{Minchin}, R.~F., {Disney}, M.~J., {Parker}, Q.~A., {et~al.} 2004, \mnras, 355,
  1303

\bibitem[{{O'Neil} \& {Bothun}(2000)}]{oneil-bothun-2000}
{O'Neil}, K. \& {Bothun}, G. 2000, \apj, 529, 811

\bibitem[{{O'Neil} {et~al.}(2004){O'Neil}, {Bothun}, {van Driel}, \& {Monnier
  Ragaigne}}]{oneil-2004a}
{O'Neil}, K., {Bothun}, G., {van Driel}, W., \& {Monnier Ragaigne}, D. 2004,
  \aap, 428, 823

\bibitem[{{O'Neil} {et~al.}(1997){O'Neil}, {Bothun}, \&
  {Cornell}}]{oneil-1997a}
{O'Neil}, K., {Bothun}, G.~D., \& {Cornell}, M.~E. 1997, \aj, 113, 1212

\bibitem[{{Peng} {et~al.}(2002){Peng}, {Ho}, {Impey}, \& {Rix}}]{peng-2002}
{Peng}, C.~Y., {Ho}, L.~C., {Impey}, C.~D., \& {Rix}, H. 2002, \aj, 124, 266

\bibitem[{{Phillipps} {et~al.}(1990){Phillipps}, {Davies}, \&
  {Disney}}]{phillipps-1990}
{Phillipps}, S., {Davies}, J.~I., \& {Disney}, M.~J. 1990, \mnras, 242, 235

\bibitem[{{Schneider} {et~al.}(1998){Schneider}, {Spitzak}, \&
  {Rosenberg}}]{schneider-1998}
{Schneider}, S.~E., {Spitzak}, J.~G., \& {Rosenberg}, J.~L. 1998, \apjl, 507,
  L9

\bibitem[{{Schombert} \& {Bothun}(1988)}]{schombert-1988}
{Schombert}, J.~M. \& {Bothun}, G.~D. 1988, \aj, 95, 1389

\bibitem[{{Spitzak} \& {Schneider}(1998)}]{spitzak-1998}
{Spitzak}, J.~G. \& {Schneider}, S.~E. 1998, \apjs, 119, 159

\bibitem[{{Sprayberry} {et~al.}(1998){Sprayberry}, {Zwaan}, \&
  {Briggs}}]{sprayberry-1998}
{Sprayberry}, D., {Zwaan}, M.~A., \& {Briggs}, F.~H. 1998, in ASP Conf. Ser.
  136: Galactic Halos, 121

\bibitem[{{Tully}(1988)}]{tully-1988}
{Tully}, R.~B. 1988, {Nearby galaxies catalog} (Cambridge and New York,
  Cambridge University Press, 1988, 221 p.)

\bibitem[{{Tully} {et~al.}(1998){Tully}, {Pierce}, {Huang}, {Saunders},
  {Verheijen}, \& {Witchalls}}]{tully-1998}
{Tully}, R.~B., {Pierce}, M.~J., {Huang}, J.-S., {et~al.} 1998, \aj, 115, 2264

\bibitem[{{Zwaan}(2000)}]{zwaan-phd}
{Zwaan}, M.~A. 2000, Ph.D.~Thesis

\bibitem[{{Zwaan} {et~al.}(1997){Zwaan}, {Briggs}, {Sprayberry}, \&
  {Sorar}}]{zwaan-1997}
{Zwaan}, M.~A., {Briggs}, F.~H., {Sprayberry}, D., \& {Sorar}, E. 1997, \apj,
  490, 173

\end{thebibliography}
\end{document}